\documentclass[prd,showpacs,showkeys,preprint]{revtex4}

\usepackage{amsmath}
\usepackage{amssymb}
\usepackage{yfonts}[1988/10/03]
\usepackage{graphicx}
\newcommand {\beq}{\begin{equation}}
\newcommand {\eeq}{\end{equation}}
\newcommand {\beqa}{\begin{eqnarray}}
\newcommand {\eeqa}{\end{eqnarray}}

\begin{document}
\title{Tensor Modes’ Damping  in  Matter and Vacuum Dominated Era}
\author{  Jafar Khodagholizadeh $^{1}$\footnote{gholizadeh@ipm.ir}$^{3}$, Amir H. Abbassi$^{2}$\footnote{ahabbasi@modares.ac.ir}and Ali A. Asgari$^{2}$\footnote{aliakbar.asgari@modares.ac.ir}}
\affiliation{$^{1}$ School of physics, Institute for research in fundamental sciences (IPM), Tehran, Iran.\\
$^{2}$ Physics Department, School of sciences,Tarbiat Modares University, P.O. Box 14155-4838, Tehran, Iran \\
$^{3}$ Farhangian University, P. O. Box 11876-13311, Tehran, Iran}

\date{\today} 

\begin{abstract}
The present paper has developed an integro-differential equation to propagate cosmological gravitation waves in matter-dominated era in accounting for the presence of free streaming neutrinos as a traceless transverse tensor part of the anisotropic stress tensor. Its focus is on short and long wavelengths of GWs that enter the horizon in matter-dominated era. Results show that the anisotropic stress reduces the squared amplitude by $ 0.03\%$ for wavelengths, entering the horizon during matter-dominated phase. This reduction is less for those wavelengths that enter the horizon at $ \Lambda $ dominated era in flat spacetime. All of the calculations have been done in closed spacetime and the results have been compared with the radiation-dominated case for both flat and closed spacetimes. Finally the paper investigates the effect of closed background on the amplitude of the gravitational waves.
\end{abstract}
\pacs{98.80.Cq, 04.30.NK}
\maketitle
\section{Introduction}
Gravitational waves are generic prediction of inflation early in the universe \cite{Guth, Linde, Albrecht}. Relic GWs can affect the CMB, causing magnetic type of CMB polarization;, therefore, it is needed to know what happens to the tensor perturbation between the time of inflation and the $ \Lambda- $ dominated era, assuming that the universe is a perfect fluid \cite{Lifshitz}. S. Weinberg derived an integro-differential equation to propagate cosmological gravitation waves in terms of $ y=\dfrac{a}{a_{EQ}} $ ($ EQ $ is the era when matter and radiation density are equal). He also investigated the neutrino effects on the GWs in special radiation, dominated in flat spacetime \cite{damping}. According to the Planck report our universe appears to be spatially flat to an accuracy of $ 0.5 \% $, but in no way do the data rule out the case of $ K=1 $\cite{ Planck}. Hence in addition to the flat case, by choosing the maximally extended de-Sitter metric ($ K=1 $) as unperturbed background, it has been found that the tensor modes fluctuations (or GWs) are equal in curved spacetime within accelerated universe expansion \cite{gholizadeh}. Then the effect of neutrinos on GWs’ background have been taken into consideration as anisotropic inertia tensor \cite{gholizadeh2}. Here, similar to the S. Weinberg's work in radiation dominated, we have applied this formula on matter and $ \Lambda-$ dominated era, where anisotropic inertia tensor is determined by neutrinos and antineutrinos from the temperature ~$ \approx 10^{10}K $ when electron positron annihilation is substantially filfilled and neutrinos have decoupled from matter and radiation.
The current paper is organized as follows: Section II discusses short wavelengths, re-entering the horizon during the matter-dominated phase, for these modes are time-independent when leaving the horizon. Section III presents the GW's equation in vacuum era, discussing the effects of free streaming neutrinos on it. And in the final section the conclusion is presented along with relevant discussions and comparisons of the results from flat\cite{Stefanek} as well as closed cosmology in $ \Lambda- $ dominated era.

\section{General Short Wavelengths}

 The tensor fluctuation in curved spacetime satisfies\cite{gholizadeh}
\begin{eqnarray}\label{4}
\nabla^{2}D_{ij}-a^{2}\ddot{D}_{ij}-3a\dot{a}D_{ij}-2KD_{ij}=16\pi G\Pi_{ij}
\end{eqnarray}
Here, the dots stand for derivative respect to the ordinary time, $ K $ is curvature constant, and $ \Pi_{ij} $ is anisotropic inertia tensor, containing the photons and neutrinos, though the former makes an insignificant contribution, due to a short mean free time. As shown in Appendix A, the above equation becomes an integro-differential equation
\begin{eqnarray}\label{7}
\ddot{D}_{q}(t)+3\dfrac{\dot{a}}{a}D_{q}(t)+\dfrac{q^{2}}{a^{2}(t)}D_{q}(t)=-64\pi G\bar{\rho}_{\nu}(t)\int_{0}^{t} K(-q \int_{t^{'}}^{t}\frac{dt^{''}}{a(t^{''})})\dot{D}_{q}(t^{'})dt^{'}
\end{eqnarray}
for the flat background $ (K=0) $ and for closed spacetime is (see Appendix A)
\begin{eqnarray}\label{88}
\ddot{D}_{n}(t)+3\dfrac{\dot{a}}{a}D_{n}(t)+\dfrac{q^{2}}{a^{2}(t)}D_{n}(t)=-64\pi G\bar{\rho}(\tau)e^{-2i\tau^{2} q^{2}}\dfrac{16}{\pi q} \dfrac{\sin \tau q}{\tau^{3} q^{3}} \int_{0}^{\tau q} d\tau^{\prime}[\dot{D}_{q}(\tau^{\prime})]\dfrac{\sin \tau^{'}q}{\tau^{'}q} e^{2i\tau^{\prime 2} q^{2}}\nonumber\\
\end{eqnarray}
At first by neglecting the anisotropic inertia tensor $ \Pi_{ij}^{T} $,  the field equation (\ref{4}), governing the Fourier components of the tensor components of $ D_{ij} $, becomes
\begin{equation}\label{6}
\ddot{D}_{q}(t)+3\frac{\dot{a}}{a}\dot{D}_{q}(t)+\frac{q^{2}}{a^{2}}D_{q}(t)=0
\end{equation}
The wave number $ q $ is a continuous number, yet in closed spacetime $q^{2}=n^{2}-2 $ and $ n $ becomes a discrete one. To study the treat of $D_{q}(t) $ evolution, it is convenient to change the independent variable $ t $ to $X \equiv \frac{\bar{\rho}_{\Lambda}}{\bar{\rho}_{M}}=\frac{\bar{\rho}_{\Lambda,EQ}}{\bar{\rho}_{M,EQ}}\dfrac{a^{3}}{a_{EQ}^{3}} $, where $ a_{EQ} $,$\bar{\rho}_{\Lambda,EQ} $ and $ \bar{\rho}_{M,EQ} $ are the value of the Robertson-Walker scale factor, the energy density of vacuum, and the energy density of matter at equality of matter-$ \Lambda $ era respectively. From the Friedmann equation we will have
\begin{eqnarray}\label{10}
H_{EQ}\dfrac{dt}{\sqrt{2}}=\dfrac{d X}{3\sqrt{X+2\Omega_{K,EQ}X^{4/3}+2X^{2}}}
\end{eqnarray}
where $ H_{EQ} $ is the expansion rate and $ \Omega_{K,EQ} $, the curvature energy when matter and $ \Lambda $ densities are equal. In particular we can define $ \chi(u) $ as
\begin{eqnarray}
D_{ij}(u)=D_{ij}(0)\chi(u)
\end{eqnarray}
where $ u $ is the conformal time multiplied by the wave number $ q $ :
\begin{eqnarray}
u=q\int\limits_{0}^{t}\dfrac{dt^{'}}{a(t^{'})}
\end{eqnarray}
when the universe was still matter-dominated, $ \chi(u) $ satisfied an integro-differential equation for short wavelengths entering the horizon
\begin{equation}\label{3}
u^{2}\chi^{''}(u)+4u\chi^{'}(u)+u^{2}=-24f_{0}(\nu)\int_{0}^{u}dU K(u-U)\chi^{'}(u)
\end{equation}
where $ f_{\nu}(0)=\frac{\Omega_{\nu}}{\Omega_{\nu}+\Omega_{\gamma}}=0.40523 $ is the fraction of the energy density in neutrinos and $ K(u) $
\begin{eqnarray}\label{13}
 K(u)&\equiv& \dfrac{1}{16}\int_{-1}^{+1}dy (1-y^{2})^{2}e^{iuy}=-\dfrac{\sin u}{u^{3}}-3\dfrac{\cos u}{u^{4}}+3\dfrac{\sin u}{u^{5}}\nonumber\\ &=& \dfrac{1}{15}(j_{0}(u)+\dfrac{10}{7}j_{2}(u)+\dfrac{3}{7}j_{4}(u))
\end{eqnarray}
in which $ j_{n} $ is the spherical Bessel function. The initial conditions are
\begin{eqnarray}\label{14}
\chi(0)=1~~~~~,~~~~~\chi^{'}(0)=0
\end{eqnarray}
The homogeneous solution of Eq.(\ref{3}) is $ \dfrac{\sin u}{u^{2}} $ where $ u=\frac{3qt}{a(t)} $ and $ a(t) $ is the scale factor in matter dominated era. In the presence of the neutrinos the solution will be suppressed and the solution of the Eq.(\ref{3}) for $ u\gg 1 $ approaches
\begin{eqnarray}
\chi(u)\longrightarrow A \dfrac{\sin u}{u^{2}}
\end{eqnarray}
A solution can be given for Eq.(\ref{3}) as the series of the spherical Bessel function:
\begin{eqnarray}\label{30}
\chi(u)=\sum_{n=0} a_{n}\dfrac{j_{n}(u)}{u}
\end{eqnarray}
Inserting the Eq.(\ref{30}) in the left-hand side of the Eq.(\ref{3}), we get
\begin{eqnarray}
\sum_{n=2} (n-1)(n+2)a_{n}\dfrac{j_{n}(u)}{u}
\end{eqnarray}
Using the $ \dfrac{j_{n}(u)}{u}=\dfrac{1}{2n+1}(j_{n-1}(u)+j_{n+1}(u)) $ the above relation will become
\begin{eqnarray}
\sum_{n=1} \{\dfrac{(n-2)(n+1)}{2n-1}a_{n-1}+\dfrac{n(n+3)}{2n+3}a_{n+1}\} j_{n}(u)
\end{eqnarray}
Also the derivative of the $ \chi(u) $ is given by
\begin{eqnarray}
\chi^{'}(u)=\sum \dfrac{a_{n}}{2n+1} \{ \dfrac{1}{2n-1}[(n-1)j_{n-2}(u)-nj_{n}(u)]+\dfrac{1}{2n+3}[(n+1)j_{n}(u)-(n+2)j_{n+2}(u)]\}\nonumber\\
\end{eqnarray}
The right hand side of the Eq.(\ref{3}) can be written as $ 1.6 f_{0}(\nu) I(u) $ and $I(u)$ given by
\begin{eqnarray}\label{8}
I(u)= \sum_{m=0,2,4}  \dfrac{1}{15} d_{m} \sum_{n=0}\{ \dfrac{a_{n}}{(2n+1)(2n-1)}I_{nm}^{(1)}(u)+\dfrac{a_{n}}{(2n+1)(2n+3)}I_{nm}^{(2)}(u)\}
\end{eqnarray}
where $ I_{nm}^{(1)}(u)=\int\limits_{0}^{u} dU j_{m}(u-U)[(n-1)j_{n-2}(U)-nj_{n}(U)] $ and $I_{nm}^{(2)}(u)=\int\limits_{0}^{u} dU j_{m}(u-U)[(n+1)j_{n}(U)-(n+2)nj_{n+2}(U)]$
and $ d_{m} $ are obtained from Eq.(\ref{13}).
The contributions to the coefficient of each $j_{l}(u)$ can be evaluated straightforwardly in Appendix B. Then by replacing
\begin{eqnarray}
\sum_{m=0,2,4} \dfrac{1}{15}   \dfrac{d_{m}}{(2n+1)}(\frac{I_{nm}^{(1)}(u)}{2n-1}+\dfrac{I_{nm}^{(2)}(u)}{(2n+3)})
\end{eqnarray}
with
\begin{eqnarray}
\sum_{l,n}C_{n,l}j_{2l}(u)
\end{eqnarray}
$C_{n,l}$ become known coefficient numbers, given in Appendix B. Eventually, Eq.(\ref{3}) will be
\begin{eqnarray} \label{27}
\sum_{n=1} \{\dfrac{(n-2)(n+1)}{2n-1}a_{n-1}+\dfrac{n(n+3)}{2n+3}a_{n+1}\} j_{n}(u)=\sum_{l=0,n=1}C_{n,l}a_{2n-1}j_{2l}(u)
\end{eqnarray}
From the conditions (\ref{14}) it is known that $a_{0}=a_{2}=0$ and $a_{1}=1$ so that
all nonzero $ a_{n}'s $ could be found and with the exception of $a_{1}$, the other coefficients are observed to be quite small:
\begin{eqnarray}
a_{1}&=&1~~~,~~~a_{3}=-1.491 \times 10^{-2}~~,~~a_{5}=3.22 \times 10^{-3}~~,~~ \nonumber\\a_{7}&=&-2.88 \times 10^{-3}~~,~~a_{9}=2.44 \times 10^{-3}~~,~~a_{11}=-1.83 \times 10^{-3}
\end{eqnarray}
All of the nonzero odd order of Bessel function  go as $A$
\begin{eqnarray}
A=\sum_{n=1}^{6}a_{2n-1}=0.98600
\end{eqnarray}
So that $\chi(u)$ approaches $0.986 \dfrac{\sin u}{u^{2}}$ or $ 0.986 j_{1}(u) $ for nonzero $f_{0}(\nu)$ which is illustrated in Fig.1. As it is clear from the Fig.1, in the matter-dominated era, neutrinos have very little impact on gravitational waves when the tensor modes are deep within the horizon, whereas in radiation-dominated era the free propagation of GW's is $\dfrac{\sin u }{u} $, becoming $ 0.8024 \dfrac{\sin u}{u} $ in presence of neutrinos. Independent of neutrinos the GWs in matter dominated era as well as after ($ \Lambda -$ dominated) is weaker than the previous (i.e. radiation) dominated era. Here, we consider the background of spacetime to be curved and the tensor modes are deep inside of horizon so the integro-differential equation (\ref{7}) becomes
\begin{eqnarray}\label{101}
\dfrac{d^{2}}{d u^{2}}D_{n}(u)+(\dfrac{4}{u})\dfrac{d}{du}D_{n}(u)+D_{n}(u)=\dfrac{35.75 f_{v}(0)}{\pi}\dfrac{\sin u}{u^{5}}
\end{eqnarray}        \begin{figure}
\includegraphics[scale=0.9]{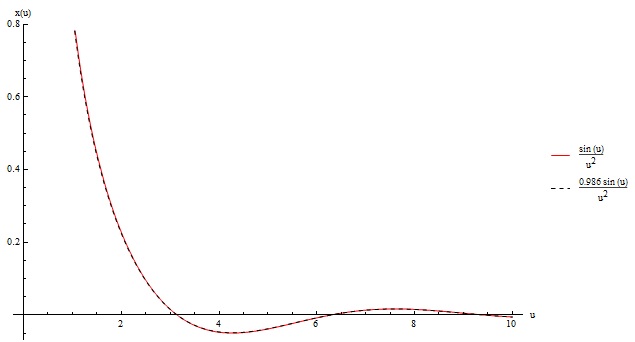}
\caption{In matter dominated era when the perturbation enters the horizon, the free-streaming neutrinos (dashed line) have no significant impact on the gravitational waves and the solution of gravitational waves equation is approximately identical to the $f_{0}(\nu)=0$ (red line), as well as in vacuum dominated era in flat spacetime.}
\label{fig:1}
\end{figure}
The general solution of above equation is
\begin{eqnarray}\label{18}
D_{n}(u)=(D_{n}^{0} -\frac{458}{75}Si(2u) + \dfrac{\frac{458}{75}Ci(2u)+\frac{229}{75}-D_{n}^{1}}{u})\dfrac{\sin u}{u^{2}}
\end{eqnarray}
where $ D_{n}^{0} $ and $ D_{n}^{1} $ are constant. $ Ci(2u) $ is the Cosine integral as $ Ci(2u) =\gamma +\ln(2u) +\int_{0}^{2u}\dfrac{\cos t -1}{t} dt $ and $Si(2u)$ is the Sine integral as $ Si(2u)= \int_{0}^{2u} \frac{\sin t}{t}$. Deep inside the horizon when $ u\gg 1 $, the right hand side of the Eq.(\ref{101}) can be entirely neglected and the solution approaches a homogeneous one as
\begin{eqnarray}
D_{n}(u)\longrightarrow \frac{\sin u}{ u^{2}}
\end{eqnarray}
 for large $ u ~(u\gg 1)$,  $ Ci(2u) $ and  $ Si(2u) $ tend to zero and $\frac{\pi}{2}$ respectively.   So
\begin{eqnarray}
D_{n}(u\gg)=(D_{n}^{0}-9.5874)\dfrac{\sin u}{u^{2}}+(\frac{4.7937-D_{n}^{1}}{u})\dfrac{\sin u}{u^{2}}
\end{eqnarray}
In $u\gg 1$, $ D_{n}(u)$ tends to be zero so the constant coefficients are $ D_{n}^{0}=9.5874 $ and $ D_{n}^{1}=4.7937 $ and also a numerical solution of Eq.(\ref{101}) shows that $ D_{n}(u) $ follows the $ f_{v}(0)=0 $ solution quite accurately until $ u \approx 1 $ when the perturbation enters horizon (as compared with the solution $ \dfrac{\sin u}{u^{2}} $ for $ f_{v}(0)=0 $). Thereafter, the solution (\ref{18}) rapidly approaches $ 0.6258 \dfrac{\sin u}{u^{2}} $. Thus the neutrino effect reduces the tensor amplitude by the factor of $ 0.6258 $ in closed cosmology, while in flat case the factor was $ 0.9860 $ in matter-dominated era. Hence the tensor contribution to the temperature multipole coefficient $ C_{l} $ and the whole of the $ ^{''}\verb"B-B"^{''} $ polarization multipole coefficient $ C_{lB}$ will be $ 0.03 \% $ ($ 60 \% $), less than what they would be without damping, as a result of free-streaming neutrinos in flat (closed) spacetime. Consequently, in mater-dominated era (same as the radiation-dominated one) and in closed cosmology, the amplitude of the gravitational waves at the presence of neutrinos will be less than the flat case or the neutrinos will have a greater effect on the damping of gravitational waves in closed cosmology as shown in Fig.2. Also the neutrinos reduce the amplitude to same extent as the radiation dominated era.
\begin{figure}
\includegraphics[scale=1.2]{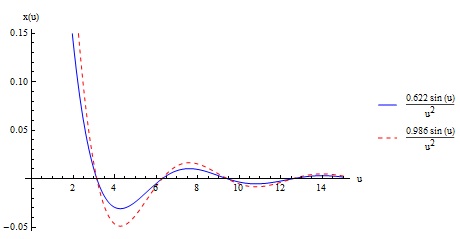}
\caption{In matter dominated era when the perturbation enters the horizon, the free-streaming neutrinos in closed cosmology (blue line) have a greater effect than the flat case (dashes) on the damping of gravitational waves and almost the same amount of amplitude reduction in radiation dominated era.}
\label{fig:2}
\end{figure}
\section{ Wavelengths at $\Lambda-$ dominated era}
To investigate the tensor perturbation may enter horizon after the $\Lambda$ density becomes important, we consider $ X\gg 1 $ so from (\ref{10}) we have
\begin{eqnarray}
H_{EQ}\dfrac{dt}{\sqrt{2}}=\dfrac{dX}{3\sqrt{X^{2}+2\Omega_{K,EQ}X^{4/3}}}
\end{eqnarray}
where $ \Omega_{K,EQ} $ is curvature density when the matter and vacuum density are equal. So then the equation (\ref{6}) becomes
\begin{eqnarray}\label{110}
(X^{2}+2\Omega_{K,EQ}X^{4/3})\dfrac{d^{2}}{dX^{2}}D_{n}(X)+(4X+\dfrac{22}{3}\Omega_{K,EQ}X^{1/3})\dfrac{d}{dX}D_{n}(X)+\dfrac{\kappa^{2}}{X^{2/3}}D_{n}(X)=0
\end{eqnarray}
where $ \kappa^{2}=\dfrac{2n^{2}}{9 H_{EQ}^{2}a_{EQ}^{2}} $ and $ n $ are discrete numbers for curved spacetime . Whatever the value of $ \kappa $ , the general solution is
\begin{eqnarray}
D_{n}(X)=\dfrac{1}{X^{4/3}} \{C_{n}^{0} Legendre P(4,\sqrt{\dfrac{32\Omega_{K,EQ}-9\kappa^{2}}{2\Omega_{K,EQ}}},\sqrt{\dfrac{\kappa^{2/3}
+2\Omega_{K,EQ}}{2\Omega_{K,EQ}}})+ \nonumber\\ C_{n}^{1}LegendreQ(4,\sqrt{\dfrac{32\Omega_{K,EQ}-9\kappa^{2}}{2\Omega_{K,EQ}}},\sqrt{\dfrac{\kappa^{2/3}
+2\Omega_{K,EQ}}{2\Omega_{K,EQ}}})\}
\end{eqnarray}
where $ C_{n}^{1} $ and $ C_{n}^{0} $ are constant. It can be seen from the above relation that the propagation of GWs today ($ \Lambda- $ dominated) can depend on the curvature energy when the matter and vacuum energy are equal. If we consider $ \Omega_{K,EQ}\ll 1 $, we can ignore it from the Eq.(\ref{110}) and then
\begin{eqnarray}
\dfrac{d^{2}}{dX^{2}}D_{n}(X)+\dfrac{4}{X}\dfrac{d}{dX}D_{n}(X)+\dfrac{\kappa^{2}}{X^{8/3}}D_{n}(X)=0
\end{eqnarray}
The solution of which is
\begin{eqnarray}
D_{n}(X)=[ (-\dfrac{\kappa}{X^{1/3}})^{4}-5(-\dfrac{\kappa}{X^{1/3}})^{2}+\dfrac{35}{27}] (C_{n}^{0} \cosh (3\sqrt{-\dfrac{\kappa}{X^{2/3}}})+ C_{n}^{1}\sinh(3\sqrt{-\dfrac{\kappa}{X^{2/3}}})) \nonumber\\
\end{eqnarray}
 The solution does not oscillate when $k\ll 1$ and in $ X\longrightarrow \infty $ separately. Then we can consider the condition $ \dfrac{\kappa}{X} \longrightarrow 0 $ tend to remain a constant value; therefore, the neutrinos have no influence on the gravitational waves. Also similar to the previous work \cite{damping,gholizadeh2}, all wavelengths take the damping value, multiplied by the neutrino, from matter-dominated era when they enter the horizon in the $\Lambda-$ dominated era as a
\begin{eqnarray}
D_{n}(X)=\alpha (k)[ (-\dfrac{\kappa}{X^{1/3}})^{4}-5(-\dfrac{\kappa}{X^{1/3}})^{2}+\dfrac{35}{27}] (C_{n}^{0} \cosh (3\sqrt{-\dfrac{\kappa}{X^{2/3}}})+ C_{n}^{1}\sinh(3\sqrt{-\dfrac{\kappa}{X^{2/3}}}))\nonumber\\
\end{eqnarray}
where $\alpha (k)$ is $ 0.9860$ for $ k\gg 1$ and $\alpha (k)=1$ for $k\ll 1$ because the damping effect is equal to zero for $k\ll 1$. Therefore it could be said $\alpha (k)\cong \dfrac{1+0.986 k}{1+k }$ and the amplitude of gravitational waves will be reduced by the factor $\alpha (k)$.
\section{The Closed Background Effective}
It is known that the $ \Omega_{K} $ is very small and, compared to $ \Omega_{\Lambda} $ and $ \Omega_{M} $, is negligible. It is not possible to consider the cases $ K=0 $ and $ K=1 $ at once, hence by taking the limit $ \Omega_{K}\longrightarrow 0 $ reach to the flat spacetimes. Here we are facing two different topologies that lead to two different structures, thus it has been shown that with the choice of the curved background, the relativistic Boltzman equation for the perturbation $ \delta n(\vec{x}, \vec{p},t) $ will change ( \ref{2} ) and also the equation of the angular distribution cosmic neutrino will alter as
\begin{eqnarray}
\dfrac{\partial}{\partial t}\Delta_{\nu}^{T}(q,\mu,t)+\dfrac{iq\mu}{a(t)}\Delta_{\nu}^{T}(q,\mu,t)-4 K\dfrac{q}{a(t)}\dfrac{\partial}{\partial \mu}\Delta_{\nu}^{T}(q,\mu,t)=-2 \dot{D}_{q}(t)~~~
\end{eqnarray}
so the neutrino contribution of the anisotropic inertia tensor $ \Pi_{ij}^{T} $ will differ from the flat background. At the end the integro-differential equation of the gravitational wave at the presence of neutrinos changes (eq.\ref{88}). Another important point is that the wave number in the closed background is discrete, which reduces the modes. Although the $ \Omega_{K} $ has no effect but as can be seen the topology is equally effective in the amplitude of the GWs at the  presence of neutrinos in radiation and matter dominated eras
\section{CONCLUSION}
We derive the propagation of gravitational waves, based on the equality parameter of matter and vacuum energy. We have seen that in the matter dominated era, the homogeneous solution of gravitational waves equation is $ \dfrac{\sin u}{u^{2}} $ whereas in earlier ( radiation dominated era) it is $\dfrac{\sin u}{u} $, so that the GWs are weaker as $ \dfrac{1}{u}$ while $ u\approx \dfrac{qt}{a(t)} $. During the matter and $\Lambda-$ dominated, the effect of the neutrinos is weak even weaker than the previous era. Also the neutrino effects in radiation dominated and flat background are weaker than the closed case as the amplitude of the GWs in flat background is $0.8026$ \cite{damping,Dicus} and for closed it is $0.4910$ \cite{gholizadeh2}. Furthermore, it has been shown that the amplitude of GWs at the presence of neutrinos in matter dominated for flat background is $0.9860$ and for closed, $0.6258$ . So then the neutrino contribution of the anisotropic inertia tensor $ \Pi_{ij}^{T} $ in closed background has approximately the same effect on the amplitude of GWs in radiation and matter era in flat spacetime, which is less at $ \Lambda- $ dominated era.
 \section*{Appendix A: Damping Effects on Wave Equation}
 We review the relativistic equation for perturbation of number density of neutrinos $ \delta n(\vec{x},\vec{p}, t)$ in closed spacetime. In general the tensor mode gravitational perturbation takes the form
\begin{eqnarray}
\delta g_{ij}=h_{ij}(\vec{x},t)=a^{2}(t) D_{ij}(\vec{x},t)
\end{eqnarray}
Where the tensor fluctuation in curved spacetime satisfies \cite{gholizadeh}
\begin{eqnarray}
\nabla^{2}D_{ij}-a^{2}\ddot{D}_{ij}-3a\dot{a}D_{ij}-2KD_{ij}=16\pi G\Pi_{ij}
\end{eqnarray}
In which $ K $ is curvature constant and $ \Pi_{ij} $, anisotropic inertia tensor. The components of the perturbed metric in Cartesian coordinate are \cite{weinberg}
\begin{eqnarray}
g_{00}=-1~,~g_{i0}=0~,~g_{ij}=a^2(t)(\delta_{ij}+K\frac{x^ix^j}{1-Kx^2}+D_{ij}(\vec{x},t))
\end{eqnarray}
 tensor perturbation  and anisotropic inertia satisfy
 \begin{eqnarray}
D_{ii}=0~~~,~~~ \nabla^{i}D_{ij}=0~~~,~~~\Pi_{ii}=0~~~,~~~\nabla^{i}\Pi_{ij}=0
\end{eqnarray}
The anisotropic inertia tensor is the sum of contributions from photons and neutrinos, though the former have an insignificant contribution to the anisotropic inertia due to a short mean free time. The latter travels without collisions when temperature drops about $ T=10^{10} K $ so neutrino distribution function in phase space has a form
\begin{equation}
n_{\nu}(\vec{x},\hat{p},t)\equiv \sum_{r}\prod_{i=1}^{3}\delta^{(3)}(x^{i}-x_{r}^{i}(t))\prod_{i=1}^{3}\delta^{(3)}(p_{i}-p_{ri}(t))
\end{equation}
Where $ r $ is individual neutrinos and anti neutrinos trajectories. In the absence of collisions terms, Boltzmann equation for neutrinos will be
\begin{eqnarray}
\dfrac{\partial n_{\nu}}{\partial t}+\dfrac{\partial n_{\nu}}{\partial x^{i}}\dot{x}^{i}+\dfrac{\partial n_{\nu}}{\partial p^{i}}p_{i}^{0}=0
\end{eqnarray}
so that $ \dot{p}_{ri}=\dfrac{1}{2p_{r}^{0}}p_{r}^{j}p_{r}^{k}(\dfrac{\partial g_{jk}}{\partial x^{i}})_{x=x_{r}} $ and $ \dot{x}_{r}^{i}=\dfrac{p_{r}^{i}}{p_{r}^{0}} $ are the change rate of momentum and the change rate of the coordinate respectively, so the above relation will be
\begin{eqnarray}\label{1}
\dfrac{\partial n_{\nu}}{\partial t}+\dfrac{\partial n_{\nu}}{\partial x^{i}}\dfrac{p^{i}}{p^{0}}+\dfrac{\partial n_{\nu}}{\partial p^{i}}\dfrac{p^{j}p^{k}}{2p^{0}}\dfrac{\partial g_{jk}}{\partial x^{i}}=0
\end{eqnarray}
$ n_{\nu}(\vec{x},t) $ in the start of free streaming has the form of the ideal gas:
\begin{eqnarray}
\bar{n}_{\nu}(\vec{x},t)=\dfrac{N_{\nu}}{(2\pi)^{3}}/[exp(\dfrac{\sqrt{g^{ij}p_{i}p_{j}}}{k_{B}a(t)\bar{T}(t)})+1]
\end{eqnarray}
 $ N_{\nu} $ is the number of neutrino types and separate antineutrinos. Moreover $ k_{B} $ is the Boltzmann constant. With a small perturbation to the metric, the neutrino distribution function gets varies a little from its equilibrium form as
\begin{eqnarray}
n_{\nu}(\vec{x},t)=n_{\nu}(a(t)\sqrt{g^{ij}p_{i}p_{j}})+\delta n_{\nu}(\vec{x},t)
\end{eqnarray}
Where $ p^{i} $ , $ p $ and $ p^{0} $ are functions of independent variable $ p_{i} $ by $p^{i}=g^{ij}p_{j}=a^{-2}(p_{i}-Kx^{i}x^{j}p_{j}) $ and $p=\sqrt{\tilde{g}^{ij} p_{i} p_{j}}=\sqrt{(\delta^{ij}-Kx^{i}x^{j})p_{i}p_{j}}$ and $ p^{0}=\sqrt{g^{ij}p_{i}p_{j}} $. Initially we assume the background spacetime to be flat, thus the first order of metric and density perturbation Eq.(\ref{1}) will be
  \begin{eqnarray}
\dfrac{\partial\delta n_{\nu}}{\partial t}+\dfrac{p^{i}}{a(t)p}\dfrac{\partial\delta n_{\nu}}{\partial x^{i}}=\dfrac{p}{2}\hat{p}_{i}\hat{p}_{j}\bar{n}_{\nu}^{\prime}(p)\dfrac{\partial}{\partial t}(a^{-2}\delta g^{ij})
\end{eqnarray}
with $ \delta g_{ij}=a^{2}D_{ij}(\vec{x},t) $, and the relativistic Boltzmann equation for the perturbation $ \delta n_{\nu}(\vec{x},\vec{p},t) $ will be
\begin{eqnarray}\label{02}
\dfrac{\partial\delta n_{\nu}}{\partial t}+\dfrac{p^{i}}{a(t)p}\dfrac{\partial\delta n_{\nu}}{\partial x^{i}}=\dfrac{p}{2}\hat{p}_{i}\hat{p}_{j}\bar{n}_{\nu}^{\prime}(p)\dfrac{\partial}{\partial t}D^{ij}(\vec{x},t)
\end{eqnarray}
We use a dimensionless intensity perturbation $ J $, defined by
\begin{eqnarray}
a^{4}(t) \bar{\rho}_{\nu}(t) J(\vec{x},\vec{p},t)\equiv N_{\nu}\int_{0}^{\infty}\delta n_{\nu}(\vec{x},\vec{p},t) 4\pi p^{3}dp
\end{eqnarray}
 and $ \bar{\rho}_{\nu} \equiv N_{\nu} a^{-4}\int 4\pi \rho^{3}\bar{n}_{\nu}(p)dp$. The Boltzmann equation (\ref{02}) becomes
\begin{eqnarray}\label{15}
\dfrac{\partial}{\partial t}J(\vec{x},\hat{p},t)+\dfrac{\hat{p}_{i}}{a(t)}\dfrac{\partial}{\partial x^{i}}J(\vec{x},\hat{p},t)=-2 \hat{p}_{i}\hat{p}_{j}\dot{D}_{ij}(\vec{x},t)
\end{eqnarray}
we will be able to find a general solution in the following form:
\begin{eqnarray}
J(\vec{x},\hat{p},t)=\sum_{\lambda=\pm 2}\int dq e^{i\vec{q}.\vec{x}}e_{ij}(\hat{q},\lambda)\beta(\hat{q},\lambda)\hat{p}_{i}\hat{p}_{j}\Delta_{\nu}^{T}(q,\hat{p}.\hat{q},t)
\end{eqnarray}
where $ \beta(\vec{q},t) $ is a stochastic parameter for the single non-decaying mode with the wave number $ q $ and the helicity $ \lambda $, $ e_{ij}(\hat{q},t) $ is the corresponding polarization tensor, and $\Delta_{\nu}^{T}(q,\hat{p}.\hat{q},t)$ is the angular distribution of cosmic neutrinos. We can define $ D_{ij}(\vec{x},t) $ as
\begin{eqnarray}
D_{ij}(\vec{x},t) =\int dq \int d^{2}\hat{q}\sum_{\lambda=\pm 2}e_{ij}(\hat{q},\lambda)\beta(\hat{q},\lambda)D_{q}(t)
\end{eqnarray}
With $ \hat{p}_{i}\hat{q}_{i}=\mu $ Eq.(\ref{15}) resulting in an equation for $ \Delta_{\nu}^{(T)} $:
\begin{eqnarray}
\dfrac{\partial}{\partial t}\Delta_{\nu}^{T}(q,\mu,t)+\dfrac{iq\mu}{a(t)}\Delta_{\nu}^{T}(q,\mu,t)=-2 \dot{D}_{q}(t)
\end{eqnarray}
A direct solution of above equation as a line of sight integral
\begin{eqnarray}
\Delta_{\nu}^{T}(q,\mu,t)=-2 \int_{t_{1}}^{t} d t^{'}\exp (-iq\mu \int_{t^{'}}^{t}\frac{dt^{''}}{a(t^{''})})\dot{D}_{q}(t^{'})
\end{eqnarray}
$\Delta_{\nu}^{T}(q,\mu,t_{1})$ is the initial value once the neutrinos are decoupled at $T=10^{10} K$ which will not be taken into consideration as the distribution of the neutrinos is essentially local thermal equilibrium and only the neutrino temperature,$\delta T_{\nu}$ and the streaming velocity,$\delta \vec{u}_{\nu}$ can perturb this distribution. What is more, these parameters do not have tensor components, thus they can be easily ignored. In the tensor mode, the only non-vanishing component is $ \delta T^{i}_{~\nu j} $:
\begin{eqnarray}
\delta T^{i}_{~\nu j}(\vec{x},t)&=&a^{-4}(t) \int d^{3}p ~\delta n_{\nu}(\vec{x},\vec{p},t)p \hat{p}_{i}\hat{p}_{j}\nonumber\\ &=&\bar{\rho}_{\nu}(t)\sum_{\lambda}  \int d^{3}q \beta (\vec{q},\lambda) e^{i\vec{q}.\vec{x}}e_{ij}(\hat{q},\lambda)\times \frac{1}{4}\int\frac{d^{2}\hat{p}}{4\pi}\Delta_{\nu}^{T}(q,\hat{p}.\hat{q},t)(1-(\hat{p}.\hat{q})^{2})^{2}\nonumber\\ &=&\sum_{\lambda}  \int d^{3}q \beta (\vec{q},\lambda) e^{i\vec{q}.\vec{x}}e_{ij}(\hat{q},\lambda) \pi_{\nu~q}^{T}(t)
\end{eqnarray}
This is the neutrino contribution of the anisotropic inertia tensor $ \Pi^{T}_{ij} $:
\begin{eqnarray}
\pi_{\nu~q}^{T}(t)&=&-4\bar{\rho}_{\nu}(t)\int\frac{d^{2}\hat{p}}{4\pi}\Delta_{\nu}^{T}(q,\hat{p}.\hat{q},t)(1-(\hat{p}.\hat{q})^{2})^{2}
\nonumber\\&=&-4\bar{\rho}_{\nu}(t)  \int_{t_{1}}^{t} d t^{'}K(q\int_{}^{}\frac{dt^{''}}{a(t^{''})})\dot{D}_{q}(t)
\end{eqnarray}
Where
\begin{eqnarray}
K(v)=\frac{j_{2}(v)}{v}=-\frac{\sin v}{v^{3}}-\frac{3\cos v}{v^{4}}+\frac{3\sin v}{v^{5}}
\end{eqnarray}
so the gravitational wave equation  now becomes an integro-differential equation:
\begin{eqnarray}\label{7}
\ddot{D}_{q}(t)+3\dfrac{\dot{a}}{a}D_{q}(t)+\dfrac{q^{2}}{a^{2}(t)}D_{q}(t)=-64\pi G\bar{\rho}_{\nu}(t)\int_{0}^{t} K(-q \int_{t^{'}}^{t}\frac{dt^{''}}{a(t^{''})})\dot{D}_{q}(t^{'})dt^{'}
\end{eqnarray}
Assuming the background is curved, with $ \delta g_{ij}=a^{2}D_{ij}(\vec{x},t) $, the relativistic Boltzmann equation for the perturbation $ \delta n_{\nu}(\vec{x},\vec{p},t) $ will be \cite{gholizadeh2}
\begin{eqnarray}\label{2}
\dfrac{\partial\delta n_{\nu}}{\partial t}+\dfrac{p^{i}}{a(t)p}\dfrac{\partial\delta n_{\nu}}{\partial x^{i}}+K \dfrac{\hat{p}_{i}}{a(t)}p x^{l}\hat{p}_{l}\dfrac{\partial\delta n_{\nu}}{\partial p_{i}}=\dfrac{p}{2}\hat{p}_{i}\hat{p}_{j}\bar{n}_{\nu}^{\prime}(p)\dfrac{\partial}{\partial t}D^{ij}(\vec{x},t) ~~~~~~~~~~~~~~~~\nonumber\\
-K \dfrac{\bar{n}_{\nu}^{\prime}(p)}{a}p \hat{p}_{k}x^{m}D^{km}(\vec{x},t)-K \dfrac{\bar{n}_{\nu}^{\prime}(p)}{a}p (x^{l}\hat{p}_{l})\hat{p}_{i} \hat{p}_{k}D^{ki}(\vec{x},t)+K^{2} \dfrac{\bar{n}_{\nu}^{\prime}(p)}{a}p x^{i}(x^{l}\hat{p}_{l})^{2}\hat{p}_{k} D^{ki}(\vec{x},t)\nonumber\\ -K^{2} \dfrac{\bar{n}_{\nu}^{\prime}(p)}{2a(t)}p \hat{p}_{i}x^{j}x^{k}(x^{l}\hat{p}_{l})^{2}\dfrac{\partial}{\partial x^{i}}D^{jk}(\vec{x},t)+K^{3} \dfrac{\bar{n}_{\nu}^{\prime}(p)}{2a} x^{i}x^{j}x^{k}(x^{l}\hat{p}_{l})^{3}\dfrac{\partial}{\partial x^{i}}D^{jk}(\vec{x},t)\nonumber\\
\end{eqnarray}
with the usual calculation the equation for $\Delta_{\nu}^{T}(q,\mu,t)$ in curved spacetime will be
\begin{eqnarray}
\dfrac{\partial}{\partial t}\Delta_{\nu}^{T}(q,\mu,t)+\dfrac{iq\mu}{a(t)}\Delta_{\nu}^{T}(q,\mu,t)-4 K\dfrac{q}{a(t)}\dfrac{\partial}{\partial \mu}\Delta_{\nu}^{T}(q,\mu,t)=-2 \dot{D}_{q}(t)~~~~~~~~~~~~
\end{eqnarray}
With the Green function method the solution is
\begin{eqnarray}
\Delta_{\nu}(q,\mu,\tau)= \dfrac{i}{2 \pi }e^{-i\tau q(\mu+ 2 \tau q)}\int _{-1}^{+1}d\mu^{\prime} \int_{0}^{\pi} d\tau^{\prime}[-2\dot{D}_{q}(\tau^{\prime})] \Theta(\tau^{\prime}q-2\tau q) e^{i\tau^{\prime} q(\mu^{\prime}+ 2\tau^{\prime}q)}\nonumber\\
\end{eqnarray}
The neutrino contribution of the anisotropic inertia tensor $ \Pi^{T}_{ij} $ is:
\begin{eqnarray}
\Pi^{T}_{ij}&=&\frac{\bar{\rho}_{\nu}(t)}{4}\int\frac{d^{2}\hat{p}}{4\pi}\Delta_{\nu}^{T}(q,\hat{p}.\hat{q},t)(1-(\hat{p}.\hat{q})^{2})^{2}\nonumber\\&=&-\dfrac{\bar{\rho}_{\nu}(t)}{4}(\dfrac{1}{2\pi })e^{-2i\tau^{2} q^{2}} \dfrac{1}{\tau^{5} q^{5}}[(-16\tau^{2} q^{2}+48)\sin(\tau q)-48\tau q \cos(\tau q)] \nonumber\\&\times & \int _{-1}^{+1}d\mu^{\prime} \int_{0}^{\tau q} d\tau^{\prime}[-2\dot{D}_{q}(\tau^{\prime})] e^{i\tau^{\prime} q(\mu^{\prime}+ 2\tau^{\prime}q)}
\end{eqnarray}
Finally, the integro-differential equation of gravitational waves in the presence of inertia tensor of neutrinos becomes
\begin{eqnarray}
\ddot{D}_{n}(t)+3\dfrac{\dot{a}}{a}D_{n}(t)+\dfrac{q^{2}}{a^{2}(t)}D_{n}(t)=-64\pi G\bar{\rho}(\tau)e^{-2i\tau^{2} q^{2}}\dfrac{16}{\pi q} \dfrac{\sin \tau q}{\tau^{3} q^{3}} \int_{0}^{\tau q} d\tau^{\prime}[\dot{D}_{q}(\tau^{\prime})]\dfrac{\sin \tau^{'}q}{\tau^{'}q} e^{2i\tau^{\prime 2} q^{2}}\nonumber\\
\end{eqnarray}
The next section will calculate the decay of gravitational waves in the matter dominated era for flat and closed spacetimes.
 \section*{Appendix B}
The right hand side of the Eq.(\ref{3}) is $-CI(u)$ with $C=1.6 f_{\nu}(0)=0.648368$ and $I(u)$ given by
 \begin{eqnarray}\label{8}
I(u)= \sum_{n=0} a_{n}\sum_{m=0,2,4}  \dfrac{1}{15} \frac{d_{m}}{(2n+1)} \{ \dfrac{I_{nm}^{(1)}(u)}{(2n-1)}+\dfrac{I_{nm}^{(2)}(u)}{(2n+3)}\}
\end{eqnarray}
Where $I_{nm}^{(1)}(u)$ and $I_{nm}^{(2)}(u)$ are
\begin{eqnarray}\label{56}
I_{nm}^{(1)}(u)&=&\int\limits_{0}^{u} dU j_{m}(u-U)[(n-1)j_{n-2}(U)-nj_{n}(U) ]\nonumber\\
I_{nm}^{(2)}(u)&=&\int\limits_{0}^{u} dU j_{m}(u-U)[(n+1)j_{n}(U)-(n+2)nj_{n+2}(U)]
\end{eqnarray}
In order to evaluate the $I_{nm}^{(1)}(u)$ and $I_{nm}^{(2)}(u)$, the Abramowitz and Stegun handbook have been used \cite{abra}. Initially, by using the Fourier, the transformation of a Legendre polynomial becomes a spherical Bessel function $(AS.10.1.14)$
\begin{eqnarray}\label{57}
j_{n}(u)=\frac{(-i)^{n}}{2}\int_{-1}^{+1}ds ~ e^{ius}P_{n}(s)
\end{eqnarray}
Putting Eq.(\ref{57}) in Eq.(\ref{56}) results in
\begin{eqnarray}\label{31}
I_{nm}^{(1)}(u)=\frac{(-i)^{m+n-1}}{4}\int_{-1}^{+1}ds \int_{-1}^{+1}dt \frac{e^{itu}-e^{isu}}{t-s}P_{m}(s)[(n-1)P_{n-2}(t)+nP_{n}(t)]
\end{eqnarray}
As well as for $I_{nm}^{(1)}(u)$ we have
\begin{eqnarray}\label{32}
I_{nm}^{(2)}(u)=\frac{(-i)^{m+n+1}}{4}\int_{-1}^{+1}ds \int_{-1}^{+1}dt \frac{e^{itu}-e^{isu}}{t-s}P_{m}(s)[(n+1)P_{n}(t)+(n+2)P_{n+2}(t)]
\end{eqnarray}
Defining the Legendre function of the second kin, $(AS.8.83)$
\begin{eqnarray}
Q_{n}(z)=\frac{1}{2}\int_{-1}^{+1}dx\frac{1}{z-x}P_{n}(x)
\end{eqnarray}
In Eq (\ref{31}) and Eq.(\ref{32}), the exponent s in the first term and t in the second do not appear, hence we have
\begin{eqnarray}
I_{nm}^{(1)}(u)&=&\frac{(-i)^{m+n-1}}{4} \int_{-1}^{+1} dt ~e^{itu} \{ Q_{m(t)}[(n-1)P_{n-2}(t)+n P_{n}(t)]+ P_{m}(t)[(n-1)Q_{n-2}(t)+n Q_{n}(t) ] \} \nonumber\\
I_{nm}^{(2)}(u)&=&\frac{(-i)^{m+n+1}}{4}\int_{-1}^{+1} dt ~e^{itu} \{ Q_{m(t)}[(n+1)P_{n}(t)+(n+2) P_{n+2}(t)]\nonumber\\ &  &~~~~~~~~~~~~~~~~~~~~~~~~~~~~~~~~~~~~~~+P_{m}(t)[(n+1)Q_{n}(t)+(n+2) Q_{n+2}(t) ] \}
\end{eqnarray}
By replacing $e^{itu}=\sum_{l} (2l+1) i^{l} j_{l}(u)P_{l}(u)$, the expressions for $I_{nm}^{(1)}(u)$ and $I_{nm}^{(2)}(u)$ become
\begin{eqnarray}
I_{nm}^{(1)}(u)=\sum_{l}(-i)^{m+n-l-1}\dfrac{(2l+1)}{2}\{\int_{-1}^{+1}dt P_{l}(t)Q_{m}(t)[(n-1)P_{n-2}(t)+nP_{n}(t)]\nonumber\\+\int_{-1}^{+1}dt P_{l}(t)P_{m}(t)[(n-1)Q_{n-2}(t)+nQ_{n}(t)]\}
\end{eqnarray}
\begin{eqnarray}
I_{nm}^{(2)}(u)=\sum_{l}(-i)^{m+n-l+1}\dfrac{(2l+1)}{2}\{\int_{-1}^{+1}dt P_{l}(t)Q_{m}(t)[(n+1)P_{n}(t)+(n+2)P_{n+2}(t)]\nonumber\\+\int_{-1}^{+1}dt P_{l}(t)P_{m}(t)[(n+1)Q_{n}(t)+(n+2)Q_{n+2}(t)]\}
\end{eqnarray}
To simplify the above-mentioned relations we use (AS 86.19): $Q_{m}(x)=\frac{1}{2}P_{m}(x)\ln \frac{1+x}{1-x}-W_{m-1}(x)$ where $W_{m-1}(x)=\sum_{k=0}^{\frac{m-1}{2}}\frac{2m-4k-1}{(2k+1)(m-k)}P_{m-2k-1}(x)$ and the formula $P_{l}(x)P_{m}(x)=\sum_{L=|l-m|}^{l+m}|<l,0,m,0 |L,0>|^{2}P_{L}(x)$. Also by using $P_{l}(x) Q_{m}(x)$ in term of $Q_{L}(x)$'s as $P_{l}(x) Q_{m}(x)=\sum_{L=|l-m|}^{l+m}[|<l,0,m,0 |L,0>|^{2}(Q_{L}(x)+W_{L-1}(x))]-P_{l}(x)W_{m-1}(x)$. In $I_{nm}(u) $ 's , the products of $P_{n}(x)'s $ and $Q_{m}(x)'s$ cancel (AS 8.14.10):
\begin{eqnarray}
\int_{-1}^{+1}dx( Q_{L}(x)P_{m\pm1}(x)+P_{L}(x)Q_{m\pm1}(x))=0
\end{eqnarray}
Finally the $I_{nm}^{(1)}(u)$ and $I_{nm}^{(2)}(u)$ reduce to
\begin{eqnarray}\label{67}
I_{nm}^{(1)}(u)&=&\sum_{l}(-i)^{m+n-l-1}\frac{2l+1}{2}j_{l}(u)\{\int_{-1}^{+1}dt \sum_{L=|l-m|}^{l+m}|<l,0,m,0 |L,0>|^{2}\nonumber\\ && \times W_{L-1}(t)[(n-1)P_{n-2}(t)+nP_{n}(t)]-\int_{-1}^{+1}dt P_{l}(t)W_{m-1}(t)[(n-1)P_{n-2}(t)+nP_{n}(t)]\}\nonumber\\
\end{eqnarray}
\begin{eqnarray}\label{68}
I_{nm}^{(2)}(u)&=&\sum_{l}(-i)^{m+n-l+1}\frac{2l+1}{2}j_{l}(u)\{\int_{-1}^{+1}dt \sum_{L=|l-m|}^{l+m}|<l,0,m,0 |L,0>|^{2}W_{L-1}(t)\nonumber\\ && \times [(n+1)P_{n}(t)+(n+2)P_{n+2}(t)]-\int_{-1}^{+1}dt P_{l}(t)W_{m-1}(t)[(n+1)P_{n}(t)+(n+2)P_{n+2}(t)]\}\nonumber\\
\end{eqnarray}
The contributions of the coefficient of each $j_{l}(u)$ in $I_{nm}^{(1)}(u)$ and $I_{nm}^{(2)}(u)$ can be evaluated. The sum over $l$ in $I_{nm}(u)'s$ is the sum of the contributions from the three terms in the kernel $m=0,2,4$. For example for $m=0$ the $I_{nm}^{(1)}(u)$ and $I_{nm}^{(2)}(u)$ will be
\begin{eqnarray}
I_{nm}^{(1)}(u)=\sum_{l}(-i)^{n-l-1}\frac{2l+1}{2}|<l,0,0,0 |L,0>|^{2}j_{l}(u)\{ \frac{4(n-1)}{(l-n+2)(l+n-1)}+\frac{4n}{(l-n)(l+n+1)}\}\nonumber\\
\end{eqnarray}
and
\begin{eqnarray}
I_{nm}^{(2)}(u)=\sum_{l}(-i)^{n-l+1}\frac{2l+1}{2}|<l,0,0,0 |L,0>|^{2}j_{l}(u)\{ \frac{4(n+1)}{(l-n)(l+n+1)}+\frac{4(n+2)}{(l-n-2)(l+n+3)}\}\nonumber\\
\end{eqnarray}
For $m=2,4$ the expressions are very big and complicated. Therefore by using the numerical methods we determine the nonzero coefficients. The Eq(\ref{3}) will be
\begin{eqnarray} \label{27}
\sum_{n=1} \{\dfrac{(n-2)(n+1)}{2n-1}a_{n-1}+\dfrac{n(n+3)}{2n+3}a_{2n-1}\} j_{n}(u)=-C\sum_{l=0,n=1}C_{n,l}a_{2n+1}j_{2l}(u)
\end{eqnarray}
So that it can be written $ C_{n,l} $ based on the $I_{nm}^{(1)}(u)$ and $I_{nm}^{(2)}(u)$. Consequently from the above equation, (\ref{67}) and (\ref{68}) we have
\begin{eqnarray}
\frac{10}{7}a_{3}&=&d_{0}\frac{a_{1}}{3}\frac{I_{10}^{(2)}(u)}{5}+ d_{2}[\frac{a_{1}}{3}(I_{12}^{(1)}(u)+\frac{I_{12}^{(2)}(u)}{5})+\frac{a_{3}}{7}\frac{I_{32}^{(1)}(u)}{5}]\nonumber\\ &&~~~~~~~~~~~~~~~+d_{4}
[\frac{a_{1}}{3}I_{14}^{(1)}(u)+\frac{a_{3}}{7}\frac{I_{34}^{(1)}(u)}{5}+\frac{a_{1}}{3}\frac{I_{14}^{(2)}(u)}{5}]\nonumber\\
\frac{10}{7}a_{3}+\frac{28}{11}a_{5}&=&d_{0}\frac{a_{1}}{3}\frac{I_{10}^{(2)}(u)}{5}+ d_{2}[\frac{a_{1}}{3}(I_{12}^{(1)}(u)+\frac{I_{12}^{(2)}(u)}{5})+\frac{a_{3}}{7}\frac{I_{32}^{(1)}(u)}{5}]\nonumber\\ &&~~~~~~~~~~~~~~~+d_{4}
[\frac{a_{1}}{3}(I_{14}^{(1)}(u)+\frac{I_{14}^{(2)}(u)}{5})+\frac{a_{3}}{7}\frac{I_{34}^{(1)}(u)}{5}]\nonumber\\
\frac{28}{11}a_{5}+\frac{54}{15}a_{7}&=&d_{0}\frac{a_{1}}{3}\frac{I_{10}^{(2)}(u)}{5}+ d_{2}[\frac{a_{3}}{7}\frac{I_{32}^{(1)}(u)}{5}+\frac{a_{1}}{3}\frac{I_{12}^{(2)}(u)}{5})]+d_{4}
[\frac{a_{3}}{7}\frac{I_{34}^{(1)}(u)}{5}+\frac{a_{1}}{3}\frac{I_{14}^{(2)}(u)}{5}]\nonumber\\
\frac{54}{15}a_{7}+\frac{88}{19}a_{9}&=&d_{0}\frac{a_{1}}{3}\frac{I_{10}^{(2)}(u)}{5}\nonumber\\
\frac{88}{19}a_{9}+\frac{130}{23}a_{11}&=&d_{0}\frac{a_{1}}{3}\frac{I_{10}^{(2)}(u)}{5}\nonumber\\
\end{eqnarray}
And other $I_{nm}^{(1)}(u)$ and $I_{nm}^{(2)}(u)$ are equal to zero. With $a_{1}=1$ and $a_{0}= a_{2}=0 $, n even all vanish and we find the $a_{2n-1}$'s decrease easily:
 \begin{eqnarray}
a_{1}=1~~~~~,~~~~~a_{3}=-1.4\times 10^{-2}~~~~~,~~~~~a_{5}=3.22\times 10^{-3}~~~~~,~~~~~ \nonumber\\
a_{7}=-2.88\times 10^{-3}~~~~~,~~~~~a_{9}=2.44\times 10^{-3}~~~~~,~~~~~a_{11}=-1.88\times 10^{-3}
\end{eqnarray}
Therefore all of the nonzero odd order Bessel function  go as $A$
\begin{eqnarray}
A=\sum_{n=1}^{6}a_{2n-1}=0.98600
\end{eqnarray}
\bibliography{apssamp}

\end{document}